\documentclass[prd,superscriptaddress,twocolumn,eqsecnum,showpacs]{revtex4}

\usepackage[dvips]{graphicx}
\usepackage{color}
\usepackage{latexsym}
\usepackage{amsmath}
\usepackage{amssymb}

\newcommand{\red}[1]{\textcolor{black}{#1}}

\newcommand{\vier}[2]{\mathrm{V}^{#1}_{\phantom{#1}#2}}
\newcommand{\vinv}[2]{\mathrm{\bf{V}}^{#2}_{\phantom{#2}#1}}

\newcommand{\vvier}[2]{\bar {\mathrm{V}}^{#1}_{\phantom{#1}#2}}
\newcommand{\vvinv}[2]{\bar {\mathrm{\bf{V}}}^{#2}_{\phantom{#2}#1}}

\begin{document}
%
%
%
\preprint{LA-UR-06-3997}
\title[Deq]
   {On the forward cone quantization of the Dirac field in
   ``longitudinal boost-invariant'' coordinates with cylindrical symmetry}

\author{Bogdan Mihaila}
\email{bmihaila@lanl.gov}
\affiliation{Materials Science and Technology Division,
   Los Alamos National Laboratory,
   Los Alamos, NM 87545}

\author{John F. Dawson}
\email{john.dawson@unh.edu}
\affiliation{
   Department of Physics,
   University of New Hampshire,
   Durham, NH 03824}

\author{Fred Cooper}
\email{cooper@santafe.edu}
\affiliation{Santa Fe Institute,
   Santa Fe, NM 87501}
\affiliation{Theoretical Division,
   Los Alamos National Laboratory,
   Los Alamos, NM 87545}


\begin{abstract}
   We obtain a complete set of free-field solutions of the Dirac equation in a
   (longitudinal) boost-invariant geometry with azimuthal symmetry
   and use these solutions to perform the canonical quantization of a free Dirac
   field of mass $M$. This coordinate system which uses the
   1+1 dimensional fluid rapidity $\eta = 1/2 \ln [(t-z)/(t+z)]$ and the fluid
   proper time $\tau = (t^2-z^2)^{1/2}$ is relevant for understanding particle
   production of quarks and antiquarks following an ultrarelativistic collision
   of heavy ions, as it incorporates the (approximate) longitudinal ``boost invariance''
   of the distribution of outgoing particles. We compare two approaches to solving
   the Dirac equation in curvilinear coordinates, one directly using Vierbeins,
   and one using a ``diagonal'' Vierbein representation.
\end{abstract}


\maketitle


\section{Introduction}
\label{sec:intro}

The kinematics of heavy-ion collisions at relativistic colliders is
such that the distribution of secondary particles obeys an
approximate longitudinal boost invariance. This approximate
invariance was the basis for many hydrodynamical simulations of
particle production following the collision of two ultrarelativistic
heavy ions.  The use of this symmetry for heavy-ion collisions was
advocated by Bjorken~\cite{Bjorken} in a seminal paper which was
based on earlier work using Landau's hydrodynamic
model~\cite{Landau} by Cooper, Frye and Schonberg~\cite{CFS} and
Hwa~\cite{Hwa}. In a hydrodynamic description of the time evolution
of the plasma produced following a collision, this invariance leads
to the surfaces of constant energy density being functions primarily
of the kinematic variable $\tau = (t^2-z^2)^{1/2}$, where the  $z$ axis is the axis of cylindrical symmetry.
For head-on collisions of identical nuclei, the kinematics also has
cylindrical symmetry.

In trying to understand the particle production of hadrons in this
scenario from first principles, one approach is to assume that a
significant fraction of the initial energy that is eventually
transformed into the final hadrons is in a classical gluonic field
configuration which produces quarks and quantum gluons by the
Schwinger tunneling mechanism~\cite{Schwinger}. In this
approximation the particles are produced by nonperturbative
tunneling. However, we ignore production due to further quantum
corrections due to rescattering of the  quarks and gluons that are
``sparked'' from the vacuum.

In this boost-invariant situation, all physical quantities such as
energy densities are just functions of the proper time $\tau$. In
particular the chromoelectric field density ${\vec E}_\alpha$,
$\alpha=1,2, \ldots 8$ is also a function of $\tau$. Choosing the
chromoelectric field in the longitudinal direction, then we can
choose a gauge where the vector potential is also a function only of
$u = \ln \tau$  and is defined via the relationship
\begin{equation}
E_\alpha^3(u) = - \frac {\partial A_\alpha^3(u)}{\partial u}
\end{equation}
In this picture the ``classical gluonic field''  $ A_\alpha^3$
would approximately be a function of $\tau$ and one would obtain the
initial quantum plasma by studying the coupled problem of  the
produced quarks and gluons interacting with the background classical
gluonic field. Such a picture was advocated by Cooper \emph{et
al}.~\cite{CEKMS} for the production of particles in the central
rapidity region. Once such a system of equations is solved, one then
has to construct the physical quantities of QCD which depend on the
two Casimir invariants of color SU(3). It is only these quantities
which are gauge invariant.

In the hydrodynamic scaling regime where the fluid velocity $v=z/t$,
the fluid rapidity $\eta = 1/2 \ln [(t-z)/(t+z)]$ can be identified
with the particle rapidity: $1/2 \ln [(1-v)/(1+v)]$. In
Ref.~\cite{CEKMS} numerical simulations were performed in 1+1
dimensions, with an a Abelian external field. Although the formalism
for doing the 3+1 dimensional calculation was presented, simulations
were not done in that paper and the cylindrical symmetry of the
head-on collision problem was not exploited.  By exploiting the
cylindrical symmetry, the computational complexity of the back
reaction problem is reduced from $N^3$ modes to describe the quantum
field of the quarks to $N^2$ modes which is quite a savings.

Recently the problem of finding the transverse momentum distribution
of quarks and gluons produces by a strong  constant chromoelectric
field by the Schwinger mechanism~\cite{Schwinger}  has been
solved~\cite{Nayak}. This calculation  found the interesting
physical result that the transverse distributions of produced
particles depends not only on an invariant related to the energy
density of the plasma, but also an invariant which describes the
direction in color space that the initial chromoelectric field was
pointing. This might have important experimental consequences.
In this paper we will discuss the canonical quantization of the free
Dirac field in the coordinate system which utilizes the curvilinear
coordinates $\tau$ and the ``fluid'' longitudinal rapidity $\eta =
\frac{1}{2}  \ln [(t-z)/(t+z)]$ as well as the cylindrical
coordinates $\rho$ and $\phi$ that correspond to the directions
transverse to the collision axis. This formulation will simplify
identifying the adiabatic solutions to the back reaction problem and
will simplify the computational effort needed to determine
transverse momentum distributions. The canonical quantization of the
fermion field presented here will also allow us to precisely define
the initial state of the quarks before the chromoelectric field is
``turned on'', so that the initial values of the time evolution
problem and back reaction can be specified.

The problem of quantizing the Dirac field in an arbitrary coordinate
system was first discussed in the classic paper of Brill and
Wheeler~\cite{Wheeler} who were concerned with understanding the
motion of electrons and neutrinos in a background gravitational
field. This problem was also later considered by
Parker~\cite{Parker} and collaborators who were interested in the
problem of particle creation in expanding universes. More recently,
Viallalba and coworkers have concerned themselves with the general
problem of separation of variables~\cite{villalba} in external
vector fields in order to find exact solutions~\cite{exact} of the
Dirac equation. The problem of quantizing the Dirac equation on the
hyperboloids $\tau = (t^2-x^2-y^2-z^2)^{1/2}$ was also studied in
great detail in the 1970's by Sommerfield~\cite{Sommerfield},
diSessa~\cite{diSessa} and others~\cite{Gromes}. Furthermore, the
problem of quantizing the Dirac equation in cylindrical coordinates
was studied by Balantekin and DeWeerd~\cite{baha} also with interest
in ultra-relativistic heavy-ion collisions, but without exploiting
the boost invariance aspect of this problem.

The intent of this paper is to fill the apparent gap in the previous
work on this subject by studying the Dirac equation in a situation
where the 1+1 dimensional fluid proper time $\tau=(t^2-z^2)^{1/2}$
is important as well as the cylindrical symmetry.  In doing so we
will also compare and relate two different ways of considering this
problem, first directly using Vierbeins (tetrads)\cite{Weyl} to
transform into the curvilinear coordinate system, and the second
making a similarity transform into a frame where the Vierbeins are
diagonal.

The paper is organized as follows: In Sec.~\ref{sec:frame} we will
review the general mathematical framework of the subsequent
discussion and outline our approach. In Sec.~\ref{sec:2d} we will
discuss the 1+1 realization of this problem using boost-invariant
coordinates. In Sec.~\ref{sec:4dcyl} we will discuss the case of a
3+1 dimensional Dirac equation in cylindrical coordinates. Finally,
in Sec.~\ref{sec:4d} we will combine the results of the previous two
sections in order to  discuss the solution of the Dirac equation in
longitudinal boost-invariant coordinates with azimuthal symmetry. We
conclude in~Sec.~\ref{sec:concl}.


\section{General framework}
\label{sec:frame}

The free-field Dirac equation in Cartesian coordinates
($\xi^\alpha$) is given by
\begin{equation}
   \bigl( \gamma^\alpha \, \partial_\alpha
   \, + \,
   M \bigr ) \ \psi
   \ = \ 0
\label{dirac_0}
   \>.
\end{equation}
Throughout this paper we use the metric
   \begin{equation}
      \eta^{\alpha \beta}=\mathrm{diag}(-1,1,1,1)
\label{metric}
      \>,
   \end{equation}
and the Dirac gamma matrices in the chiral representation, in 3+1
dimensions, given as
\begin{align}
   \gamma^0 =
   \left [
   \begin{array}{cc}
   0 & 1 \\
   -1 & 0
   \end{array}
   \right ]
   , \quad
   \gamma^k =
   \left [
   \begin{array}{cc}
   0 & \sigma_k \\
   \sigma_k & 0
   \end{array}
   \right ]
\label{gammas}
   ,
\end{align}
where $\sigma^k$ are the usual Pauli matrices, and $1$ is the unit
($2\times 2$) matrix.  The constant Dirac matrices satisfy the
relation
\begin{equation}
   \bigl \{ \gamma^\alpha,\, \gamma^\beta \bigr \}
   \ = \
   2 \ \eta^{\alpha \beta}
   \>.
\end{equation}

Next, we will consider the change of variables
($\xi^\alpha$~$\rightarrow$~$x^a$). We introduce the Vierbeins,
$\vier{\alpha}{a}$, and their inverses, $\vinv{\alpha}{a}$, as
\begin{align}
   \mathrm{d} \xi^\alpha =
   \frac{\partial \xi^\alpha}{\partial x^a} \ \mathrm{d}x^a
   = & \
   \vier{\alpha}{a} \ \mathrm{d}x^a
   \>,
   \\
   \partial_\alpha =
   \frac{\partial}{\partial \xi^\alpha} =
    \frac{\partial x^a}{\partial \xi^\alpha}\frac{\partial}{\partial x^a} \
    = & \
   \vinv{\alpha}{a}    \partial_a
   \>.
\end{align}
We have
\begin{equation}
   \vier{\alpha}{a} \vinv{\beta}{a} =
   \delta_{\alpha \beta}
   \>,
   \quad
   \vinv{\alpha}{a} \vier{\alpha}{b} =
   \delta_{ab}
   \>.
\end{equation}
Under the change of variables the line element transforms as
\begin{align}
   \mathrm{d}s^2
   = \eta_{\alpha \beta} \,
   \mathrm{d}\xi^\alpha \mathrm{d}\xi^\beta
   = g_{ab} \,
   \mathrm{d}x^a \mathrm{d}x^b
   \>.
\end{align}
In the following we will use the notation $\sqrt{-g}=\det[g_{ab}]$.
One also has the connection between the Cartesian metric and the
metric in the curvilinear coordinates
\begin{equation}
   g_{ab} = { \mathrm{\bf V} }^{\alpha}_{\phantom{\alpha}a} \,
            { \mathrm{\bf V} }^{\beta}_{\phantom{\beta}b} \ \eta_{\alpha \beta}
   \>.
\end{equation}
Finally, in the new system of coordinates the free-field Dirac
equation~\eqref{dirac_0} becomes
\begin{equation}
   \bigl( \tilde \gamma^a \, \partial_a
   \, + \,
   M \bigr ) \ \psi
   \ = \ 0
\label{dirac_1}
   \>.
\end{equation}

To simplify the Dirac equation in an arbitrary curvilinear
coordinate system one often performs a similarity transformation to
diagonalize the Vierbein matrices. This then produces a
representation which we will connote the ``diagonal tetrad''
representation. To obtain the diagonal tetrad representation, one
introduces a similarity transformation, $S$, such that
\begin{align}
   S^{-1} \ \tilde \gamma^a \ S \ = \ \bar \gamma^a
   \>,
\end{align}
where the gamma matrices $\bar \gamma^a$ are proportional to the
constant gamma matrices, $\gamma^a$. By construction, the two set of
gamma matrices, $\tilde \gamma^a$ and $\bar \gamma^a$, are defined
such that
\begin{equation}
   \bar \gamma^\alpha
   =
   \vvinv{\alpha}{a} \,
   \gamma^\alpha
   \>,
\end{equation}
where $\vvinv{\alpha}{a}$ are the Vierbeins in the diagonal tetrad
representation. The two sets of gamma matrices, $\tilde \gamma^a$
and $\bar \gamma^a$ satisfy the same algebra
\begin{equation}
   \bigl \{ \tilde \gamma^a,\, \tilde \gamma^b \bigr \}
   \ = \
   \bigl \{ \bar \gamma^a,\, \bar \gamma^b \bigr \}
   \ = \
   2 \ g^{ab}
   \>.
\end{equation}
Hence, the similarity transformation $S$ which connects these two
irreducible sets of matrices, is unique~\cite{jauch}.

In the diagonal tetrad representation the Dirac equation becomes
\begin{equation}
   \bigl( \bar \gamma^a \, \nabla_a
   \, + \,
   M \bigr ) \ \psi_{\mathrm{[d]}}
   \ = \ 0
   \>,
\end{equation}
where we have introduced the diagonal Dirac wave function
\begin{equation}
   \psi_{\mathrm{[d]}} \ = \
   S^{-1} \ \psi
   \>,
\end{equation}
and the covariant derivatives
\begin{equation}
   \nabla_a \ = \
   S^{-1} \ \partial_a \ S
   \ = \
   \partial_a \ - \
   \Gamma_a
   \>.
\end{equation}
Here, the \emph{spinor} connection $\Gamma_a$ is given by
\begin{align}
   \Gamma_a = & \
   \frac{1}{2} \, \Sigma^{\alpha \beta} \, \eta_{\alpha \gamma}
   \vvier{\gamma}{b} \,
   \bigl ( \partial_a \vvinv{\beta}{b} + \Gamma_{ac}^b \vvinv{\beta}{c}
   \bigr )
   \>,
\end{align}
and we have introduced the notation $\Sigma^{\alpha \beta} =
\frac{1}{4} [ \gamma^\alpha, \, \gamma^\beta ]$, defined in terms of
the constant Dirac matrices.

The Christoffel symbols, $\Gamma^a_{bc}$, are obtained from the
Lagrange equations of motion
\begin{align}
   \frac{\mathrm{d}}{\mathrm{d}t}
   \Bigl ( \frac{\partial \mathcal L}{\partial \dot x^a} \Bigl )
   \ - \
   \frac{\partial \mathcal L}{\partial x^a}
   = \ 0 \>,
\end{align}
or
\begin{align}
   \ddot x^a + \Gamma^a_{bc} \, \dot x^b \dot x^c
   = \ 0 \>,
\end{align}
corresponding to the Lagrangian of a free particle in the $x^a$
coordinates~\cite{bd}, i.e.
\begin{equation}
   \mathcal{L} =
   g_{ab} \ \dot x^a \, \dot x^b
   \>.
\end{equation}

The inner product is defined by considering the covariant
conservation law for a current. In an arbitrary number ($d$) of
\emph{spatial} dimensions, we have the conservation
law~\cite{weinberg}
\begin{equation}
   \int \mathrm{d}^{d+1} x \ \sqrt{-g} \ \mathcal{V}^\mu_{\phantom{\mu};\mu}
   \ = \ 0
   \>,
\end{equation}
provided that $\mathcal{V}^\mu$ vanishes at infinity. Here the
covariant divergence is defined as
\begin{equation}
   \mathcal{V}^\mu_{\phantom{\mu};\mu}
   \ = \ \frac{1}{\sqrt{-g}} \ \partial_\mu
   \bigl ( \sqrt{-g} \ \mathcal{V}^\mu \bigr )
   \>,
\end{equation}
and the conservation law can also be written as
\begin{equation}
   \int \mathrm{d}^{d+1} x \ \partial_\mu
       \bigl ( \sqrt{-g} \ \mathcal{V}^\mu \bigr )
   \ = \ 0
   \>.
\end{equation}
or
\begin{equation}
   \int \mathrm{d} \Sigma^\mu \ \sqrt{-g} \ \mathcal{V}^\mu =0
   \>.
\end{equation}
Here $\Sigma^\mu$ is the $d$-dimensional hyper-surface perpendicular
to the $\mu$ direction. Choosing $\mu =0$ then the conserved charge
is (with $\mathcal{V}^\mu \equiv j^\mu = \bar \psi \, \tilde
\gamma^\mu \, \psi$)
\begin{equation}
   Q = \int_{x_0\mathrm{=const.}} \mathrm{d}^d x \
       \sqrt{-g} \ \bigl ( \bar \psi \, \tilde \gamma^0 \, \psi \bigr )
\label{charge}
   \>.
\end{equation}
Thus the natural Dirac inner product is :
\begin{equation}
   \bigl ( \psi_1, \, \psi_2 \bigr ) \equiv  \int_{x_0\mathrm{=const.}} \mathrm{d}^d x \ \
       \sqrt{-g} \ \bigl ( \bar \psi_1 \, \tilde \gamma^0 \, \psi_2 \bigr )
\label{inner}
   \>,
\end{equation}
with $\bar \psi \equiv \psi^\dag \gamma^0$. The canonical
quantization of the Dirac field assumes a mode expansion:
\begin{equation}
   \hat \psi(x)
   =
   \int \red{\mathrm{d}^d k} \ \Bigl [
   \psi_{\mathbf{k}}^{(+)}(x) \, a_{\mathbf{k}} +
   \psi_{\mathbf{k}}^{(-)}(x) \, b_{\mathbf{k}}^\dag
   \Bigr ]
   \>.
\end{equation}
with
\begin{equation}
   \{ a_{\mathbf{k}},\, a_{\mathbf{k}'}^\dag \}
   \ = \
   \{ b_{\mathbf{k}},\, b_{\mathbf{k}'}^\dag \}
   \ = \
   \red{
   \delta(\mathbf{k}'-\mathbf{k})
   }
   \>.
\end{equation}
Here $\psi_{k}^{(\pm)}(x)$ are positive- and negative-energy
solutions~\cite{disessa_bc} of the homogeneous Dirac equation and
satisfy the orthonormality relations
\begin{align}
   \bigl ( \bar \psi^{(\pm)}_{\mathbf{k}}, \, \psi^{(\pm)}_{\mathbf{k}'}
   \bigr )
   & =
   \red{
   \delta(\mathbf{k}' - \mathbf{k})
   }
   \>,
   \\
   \bigl ( \bar \psi^{(+)}_{\mathbf{k}}, \, \psi^{(-)}_{\mathbf{k}'}
   \bigr )
   & =
   0
   \>.
\end{align}
In the following, we will study the canonical quantization of the
free Dirac field in the Cartesian and diagonal tetrad
representations, in the context of several system of coordinates
characteristic to the physics of heavy-ion collisions.


\section{Boost-invariant coordinates in 1+1 dimensions}
\label{sec:2d}

For illustrative purposes, we consider first the 1+1 dimensional
case studied previously by diSessa~\cite{diSessa}. Here, the Dirac equation is
\begin{equation}
   \bigl (
   \gamma^0 \partial_0 + \gamma^3 \partial_3 + M
   \bigr ) \ \psi
   = 0
   .
\end{equation}
In 1+1 dimensions, the chiral representation of the Dirac gamma
matrices for our choice of metric, $\eta^{\alpha
\beta}=\mathrm{diag}(-1,1)$, is (compare with Eq.~\ref{gammas})
\begin{align}
   \gamma^0 =
   \left [
   \begin{array}{cc}
   0 & 1 \\
   -1 & 0
   \end{array}
   \right ]
   , \quad
   \gamma^3 =
   \left [
   \begin{array}{cc}
   0 & 1 \\
   1 & 0
   \end{array}
   \right ]
   .
\end{align}
Consider the change of variables $(\xi^\alpha \rightarrow x^a)$ :
\begin{align}
   t = \tau \, \cosh \eta
   \>,
   \qquad
   z = \tau \, \sinh \eta
   \>,
\end{align}
with the metric $g_{ab} = \mathrm{diag}(-1,\tau^2)$, and
$\sqrt{-g}=\tau$.


\subsection{Cartesian tetrad representation}

For our change of variables, the Vierbeins are calculated as
\begin{align*}
   \vier{\alpha}{a}
   \ = \
   \left [
   \begin{array}{cc}
   \cosh \eta & \tau \, \sinh \eta \\
   \sinh \eta & \tau \, \cosh \eta
   \end{array}
   \right ]
   \>,
\end{align*}
whereas the inverse Vierbeins are
\begin{align*}
   \vinv{\alpha}{a}
   \ = \
   \left [
   \begin{array}{cccc}
   \cosh \eta & - \sinh \eta \\
   - \displaystyle{\frac{1}{\tau}} \, \sinh \eta &
         \displaystyle{\frac{1}{\tau}} \, \cosh \eta
   \end{array}
   \right ]
   \>.
\end{align*}
We can now write the Dirac equation as
\begin{equation}
   \bigl( \tilde \gamma^a \, \partial_a
   \, + \,
   M \bigr ) \ \psi
   \ = \ 0
\label{dirac_2d}
   \>,
\end{equation}
where the Dirac matrices are given by
\begin{align}
   \tilde \gamma^0 = & \
   \gamma^0 \cosh\eta \, - \, \gamma^3 \sinh \eta
   \>,
   \\
   \tilde \gamma^3 = & \
   \frac{1}{\tau} \
   \bigl ( - \gamma^0 \sinh\eta \, + \, \gamma^3 \cosh \eta
   \bigr )
   \>.
\end{align}
Equation~\eqref{dirac_2d} can be explicitly written as
\begin{align}
   \left [
   \begin{array}{cc}
   M & \mathcal{T}_- \\
   - \mathcal{T}_+ & M
   \end{array}
   \right ] \
   \left [
   \begin{array}{r}
      \psi_1 \\
      \psi_2
   \end{array}
   \right ]
   = 0
\label{dirac_2da}
   ,
\end{align}
where we have introduced the raising and lowering operators
\begin{align}
   \mathcal{T}_- = & \
   \partial_t + \partial_z
   =
   e^{-\eta} \,
   \Bigl (
      \partial_\tau + \frac{1}{\tau} \, \partial_\eta
   \Bigr )
\label{tminus}
   \>,
   \\
   \mathcal{T}_+ = & \
   \partial_t - \partial_z
   =
   e^{\eta} \,
   \Bigl (
      \partial_\tau - \frac{1}{\tau} \, \partial_\eta
   \Bigr )
\label{tplus}
   \>,
\end{align}
which obey the properties
\begin{align}
   \mathcal{T}_- \,
   \Bigl [
      e^{\mathrm{i}k_\eta\eta} \, J_{\mathrm{i}k_\eta}(M \tau)
   \Bigr ]
   = & \ + M \,
   e^{(\mathrm{i}k_\eta-1)\eta} \, J_{\mathrm{i}k_\eta-1}(M \tau)
   \>,
   \\
   \mathcal{T}_+ \,
   \Bigl [
      e^{\mathrm{i}k_\eta\eta} \, J_{\mathrm{i}k_\eta}(M \tau)
   \Bigr ]
   = & \ - M \,
   e^{(\mathrm{i}k_\eta+1)\eta} \, J_{\mathrm{i}k_\eta+1}(M \tau)
   \>,
\end{align}
where $J_{\mathrm{i}k_\eta}(M \tau)$ are Bessel functions of
\emph{complex} order (and \emph{real}
argument)~\cite{bessel_complex}. Throughout this paper we use the
definitions and properties of the Bessel functions as presented in
Ref.~\cite{watson}.

Therefore, we find the amplitudes $\psi_1$ and $\psi_2$ obey the
equations
\begin{align}
   \bigl (
      \mathcal{T}_- \mathcal{T}_+
      \ + \
      M^2
   \bigr ) \, \psi_1
   = & \ 0
   \>,
   \\
   \bigl (
      \mathcal{T}_+ \mathcal{T}_-
      \ + \
      M^2
   \bigr ) \, \psi_2
   = & \ 0
   \>.
\end{align}
Alternatively we could have ``squared'' the Dirac
equation~\eqref{dirac_2d} in the usual textbook fashion. This would
have doubled the number of solutions and we would have kept only one
set such as done in~\cite{kluger}.


The solutions of the above equations have the general form
\begin{equation}
   \psi(\tau,\eta) =
   \int \red{\mathrm{d}k_\eta} \ \psi_{k_\eta}(\tau,\eta)
   \>.
\end{equation}
It is necessary~\cite{non_symmetric} to seek a solution
$\psi_{k_\eta}$ of the form
\begin{align}
   \psi_{k_\eta} =
   N_{k_\eta} \,
   e^{\mathrm{i}k_\eta \eta} \,
   \left [
   \begin{array}{c}
   e^{-\eta/2} \, J_{\mathrm{i}k_\eta-\frac{1}{2}}(M\tau) \\
   - \, e^{\eta/2} \, J_{\mathrm{i}k_\eta+\frac{1}{2}}(M\tau)
   \end{array}
   \right ]
\label{eq:psi_2dc}
   \>.
\end{align}
Requiring the usual positive/negative-energy boundary
conditions~\cite{disessa_bc}, we obtain
\begin{align}
   \psi_{k_\eta}^{(+)}
   = N_{k_\eta}^{(+)} \,
   e^{\mathrm{i}k_\eta \eta} \,
   \left [
   \begin{array}{c}
   e^{-\eta/2} \, H^{(2)}_{\mathrm{i}k_\eta-\frac{1}{2}}(M\tau) \\
   - \, e^{\eta/2} \, H^{(2)}_{\mathrm{i}k_\eta+\frac{1}{2}}(M\tau)
   \end{array}
   \right ]
   ,
\end{align}
and
\begin{align}
   \psi_{k_\eta}^{(-)}
   = N_{k_\eta}^{(-)} \,
   e^{\mathrm{i}k_\eta \eta} \,
   \left [
   \begin{array}{c}
   e^{-\eta/2} \, H^{(1)}_{\mathrm{i}k_\eta-\frac{1}{2}}(M\tau) \\
   - \, e^{\eta/2} \, H^{(1)}_{\mathrm{i}k_\eta+\frac{1}{2}}(M\tau)
   \end{array}
   \right ]
   .
\end{align}
Then, the canonical quantization of the Dirac field is obtained as
\begin{equation}
   \hat \psi(\tau,\eta)
   =
   \int \red{\mathrm{d}k_\eta} \ \Bigl [
   \psi_{k_\eta}^{(+)}(\tau,\eta) \, a_{k_\eta} +
   \psi_{k_\eta}^{(-)}(\tau,\eta) \, b_{k_\eta}^\dag
   \Bigr ]
   \>.
\end{equation}
The normalization constants $N_{k_\eta}^{(\pm)}$ are obtained by
requiring the fields $\psi_{k_\eta}^{(\pm)}$ to satisfy the
orthonormality relations
\begin{align}
   \bigl ( \bar \psi^{(\pm)}_{k_\eta}, \, \psi^{(\pm)}_{k_\eta'}
   \bigr )
   & =
   \red{
   \delta(k_\eta'-k_\eta)
   }
   \>,
\label{plus}
   \\
   \bigl ( \bar \psi^{(+)}_{k_\eta}, \, \psi^{(-)}_{k_\eta'}
   \bigr )
   & =
   0
   \>,
\end{align}
where the inner product~\eqref{inner} gives
\begin{align}
   \bigl ( \bar \psi_{k_\eta}, \, \psi_{k_\eta'}
   \bigr )
   = & \, - \tau
   \int \mathrm{d}\eta \
   \bar \psi_{k_\eta} \, \tilde \gamma^0 \, \psi_{k_\eta'}
\label{dotprod}
   \\ \notag
   = & \ \tau
   \int \mathrm{d}\eta \
   \Bigl [ \psi^{(+)\star}_{1,k_\eta} \, \psi^{(+)}_{1,k_\eta'} \, e^{\eta}
   +
   \psi^{(+)\star}_{2,k_\eta} \, \psi^{(+)}_{2,k_\eta'} \, e^{-\eta}
   \Bigr ]
   \>,
\end{align}
with $\bar \psi = \psi^\dag \, \gamma^0$.
We work out each term separately:
\begin{widetext}
\begin{align}
   \psi^{(+)\star}_{1,k_\eta} \, \psi^{(+)}_{1,k_\eta'} \, e^{\eta}
   = \ &
   N_{k_\eta}^{(+)\star}
   N_{k_\eta'}^{(+)} \,
   \mathrm{i} \,
   e^{\mathrm{i}(k_\eta'-k_\eta) \eta} \,
   e^{- \pi k_\eta} \,
   H^{(1)}_{\mathrm{i}k_\eta+\frac{1}{2}} \,
   H^{(2)}_{\mathrm{i}k_\eta'-\frac{1}{2}}
   \>,
   \\
   \psi^{(+)\star}_{2,k_\eta} \, \psi^{(+)}_{2,k_\eta'} \, e^{-\eta}
   = \ &
   - \,
   N_{k_\eta}^{(+)\star}
   N_{k_\eta'}^{(+)} \,
   \mathrm{i} \,
   e^{\mathrm{i}(k_\eta'-k_\eta) \eta} \,
   e^{- \pi k_\eta} \
   H^{(1)}_{\mathrm{i}k_\eta-\frac{1}{2}} \,
   H^{(2)}_{\mathrm{i}k_\eta'+\frac{1}{2}}
   \>,
   \\
   \psi^{(-)\star}_{1,k_\eta} \, \psi^{(-)}_{1,k_\eta'} \, e^{\eta}
   = \ &
   N_{k_\eta}^{(-)\star}
   N_{k_\eta'}^{(-)} \,
   (-\mathrm{i}) \,
   e^{\mathrm{i}(k_\eta'-k_\eta) \eta} \,
   e^{\pi k_\eta} \,
   H^{(2)}_{\mathrm{i}k_\eta+\frac{1}{2}} \,
   H^{(1)}_{\mathrm{i}k_\eta'-\frac{1}{2}}
   \>,
   \\
   \psi^{(-)\star}_{2,k_\eta} \, \psi^{(-)}_{2,k_\eta'} \, e^{-\eta}
   = \ &
   - \,
   N_{k_\eta}^{(-)\star}
   N_{k_\eta'}^{(-)} \,
   (-\mathrm{i}) \,
   e^{\mathrm{i}(k_\eta'-k_\eta) \eta} \,
   e^{\pi k_\eta} \
   H^{(2)}_{\mathrm{i}k_\eta-\frac{1}{2}} \,
   H^{(1)}_{\mathrm{i}k_\eta'+\frac{1}{2}}
   \>,
\end{align}
\end{widetext}
where we have used the properties of the Hankel functions of real
argument, $x$~\cite{watson}
\begin{align}
   H^{(1,2)\star}_{\nu}(x)
   = & \
   H^{(2,1)}_{\nu^\star}(x)
   \>,
   \\
   H^{(1,2)}_{-\nu}(x)
   = & \
   e^{(+,-)\mathrm{i}\pi \nu}
   H^{(1,2)}_{\nu}(x)
   \>.
\end{align}
We will also use the Wronskian identity~\cite{watson}:
\begin{align}
   H^{(1)}_{\mathrm{i}k}(x) \,
   H^{(2)}_{\mathrm{i}k-1}(x)
   -
   H^{(1)}_{\mathrm{i}k-1}(x) \,
   H^{(2)}_{\mathrm{i}k}(x)
   =
      -
   \frac{4 \, \mathrm{i}}{\pi \, x} \>.
\end{align}
We note that by performing the integral in~\eqref{dotprod} we obtain
\begin{equation}
   \int \mathrm{d}\eta \
   e^{\mathrm{i}(k'-k) \eta}
   = 2\pi \ \delta(k'-k)
   \>.
\end{equation}
Then, we find  the normalization constants must satisfy the
conditions
\begin{align}
   1
   = \ &
   N_{k_\eta}^{(\pm)\star}
   N_{k_\eta}^{(\pm)} \
   e^{\mp \pi k_\eta} \,
   \frac{8}{M}
   \>.
\end{align}
We pick the normalization constants as:
\begin{align}
   N_{k_\eta}^{(\pm)}
   = & \
   \sqrt{ \frac{M}{8} } \
   e^{ \pm \, \pi k_\eta/2 }
   \>.
\end{align}
Therefore, we find that the positive/negative energy solutions
of~\eqref{dirac_2d} are
\begin{align}
   \psi_{k_\eta}^{(+)}
   =
   \sqrt{ \frac{M}{8} } \,
   e^{ \pi k_\eta/2 }
   e^{\mathrm{i}k_\eta \eta} \,
   \left [
   \begin{array}{c}
   e^{-\eta/2} \, H^{(2)}_{\mathrm{i}k_\eta-\frac{1}{2}}(M\tau) \\
   - \, e^{\eta/2} \, H^{(2)}_{\mathrm{i}k_\eta+\frac{1}{2}}(M\tau)
   \end{array}
   \right ]
\label{psi+a}
   ,
\end{align}
and
\begin{align}
   \psi_{k_\eta}^{(-)}
   =
   \sqrt{ \frac{M}{8} } \,
   e^{ - \, \pi k_\eta/2 }
   e^{\mathrm{i}k_\eta \eta} \,
   \left [
   \begin{array}{c}
   e^{-\eta/2} \, H^{(1)}_{\mathrm{i}k_\eta-\frac{1}{2}}(M\tau) \\
   - \, e^{\eta/2} \, H^{(1)}_{\mathrm{i}k_\eta+\frac{1}{2}}(M\tau)
   \end{array}
   \right ]
\label{psi-a}
   .
\end{align}


\subsection{Diagonal tetrad representation}

We begin with Eq.~\eqref{dirac_2d} and introduce the similarity
transformation
\begin{equation}
   S_\tau
          \ = \ \cosh \frac{\eta}{2}
                \ - \ \gamma^0 \gamma^3 \, \sinh \frac{\eta}{2}
          \ = \ \exp \Bigl ( - \frac{\eta}{2} \ \gamma^0 \gamma^3
                     \Bigr )
\label{stau}
   \>,
\end{equation}
such that
\begin{align}
   S_\tau^{-1} \ \tilde \gamma^a \ S_\tau \ = \ \bar \gamma^a
\label{gamma_transf}
   \>,
\end{align}
with
\begin{align}
   \bar \gamma^0 = \gamma^0 \>,
   \quad
   \bar \gamma^3 = \frac {1}{\tau} \ \gamma^3 \>.
\end{align}
The diagonal (rotating) Vierbein representation is identified as
\begin{align}
   \vvier{\alpha}{a} \ = \ & \mathrm{diag}(1,\tau)
   \>,
   \\
   \vvinv{\alpha}{a} \ = \ &
   \mathrm{diag}\bigl ( 1,\frac{1}{\tau} \bigr)
   \>.
\end{align}
The Dirac equation in the diagonal tetrad representation is
\begin{equation}
   \bigl( \bar \gamma^a \, \nabla_a
   \, + \,
   M \bigr ) \ \psi_{\mathrm{[d]}}
   \ = \ 0
   \>,
\end{equation}
where we have introduced the diagonal Dirac wavefunction
\begin{equation}
   \psi_{\mathrm{[d]}} \ = \
   S_\tau^{-1} \ \psi
\label{psi_tau}
   \>.
\end{equation}

The identification process described in Sec.~\ref{sec:frame} leads
to the nonzero Christoffel symbols
\begin{align}
   \Gamma^\tau_{\eta \eta} = & \ \tau \>,
   \\
   \Gamma^\eta_{\tau \eta} = \Gamma^\eta_{\eta \tau} = & \
   \frac{1}{\tau}
   \>,
\end{align}
the spinor connections
\begin{align}
   \Gamma_\eta = & \frac{1}{2} \, \gamma^0 \gamma^3
   \>,
   \quad
   \Gamma_\tau =
   0
   \>,
\end{align}
and the covariant derivatives
\begin{align}
   \nabla_\tau = \partial_\tau \>,
   \quad
   \nabla_\eta = \partial_\eta - \frac{1}{2} \, \gamma^0 \gamma^3
   \>.
\end{align}
Finally, the Dirac equation becomes
\begin{align}
   \Bigl [
     \gamma^0 \, \Bigl ( \partial_\tau + \frac{1}{2\tau} \Bigr )
   + \gamma^3 \, \frac{1}{\tau} \ \partial_\eta
   + M
   \Bigr ] \, \psi_{\mathrm{[d]}} = 0
\label{dirac_2db}
   \>,
\end{align}
and has a solution of the form
\begin{align}
   \psi_{k_\eta\mathrm{[d]}} =
   N_{k_\eta} \,
   e^{\mathrm{i}k_\eta \eta} \,
   \left [
   \begin{array}{c}
   J_{\mathrm{i}k_\eta-\frac{1}{2}}(M\tau) \\
   - \, J_{\mathrm{i}k_\eta+\frac{1}{2}}(M\tau)
   \end{array}
   \right ]
   \>.
\end{align}
We find positive- and negative-energy solutions obeying the
relations~\eqref{plus}, as
\begin{align}
   \psi_{k_\eta\mathrm{[d]}}^{(+)}
   =
   \sqrt{ \frac{M}{8} } \,
   e^{ \pi k_\eta/2 }
   e^{\mathrm{i}k_\eta \eta} \,
   \left [
   \begin{array}{c}
   H^{(2)}_{\mathrm{i}k_\eta-\frac{1}{2}}(M\tau) \\
   - \, H^{(2)}_{\mathrm{i}k_\eta+\frac{1}{2}}(M\tau)
   \end{array}
   \right ]
\label{psi+b}
   ,
\end{align}
and
\begin{align}
   \psi_{k_\eta\mathrm{[d]}}^{(-)}
   =
   \sqrt{ \frac{M}{8} } \,
   e^{ - \, \pi k_\eta/2 }
   e^{\mathrm{i}k_\eta \eta} \,
   \left [
   \begin{array}{c}
   H^{(1)}_{\mathrm{i}k_\eta-\frac{1}{2}}(M\tau) \\
   - \, H^{(1)}_{\mathrm{i}k_\eta+\frac{1}{2}}(M\tau)
   \end{array}
   \right ]
\label{psi-b}
   .
\end{align}
Finally, the Dirac fields in the Cartesian and diagonal tetrad
representations are related via the transformation~\eqref{psi_tau}.
Indeed, with our choice of gamma matrices, we obtain
\begin{align}
   \gamma^0 \gamma^3 =
   \left [
   \begin{array}{cc}
   1 & 0 \\
   0 & -1
   \end{array}
   \right ]
   ,
\end{align}
which gives the explicit form of the similarity
transformation~\eqref{stau}
\begin{align}
   S_\tau =
   \left [
   \begin{array}{cc}
   e^{-\eta/2} & 0 \\
   0 & e^{\eta/2}
   \end{array}
   \right ]
   .
\end{align}
We immediately verify that the Dirac solutions~\eqref{psi+a}
and~\eqref{psi-a} are obtained from~\eqref{psi+b} and~\eqref{psi-b}
as
\begin{equation}
   \psi^{(\pm)}_{k_\eta} = S_\tau \, \psi^{(\pm)}_{k_\eta\mathrm{[d]}}
   \>.
\end{equation}


\section{Cylindrical coordinates in 3+1 dimensions}
\label{sec:4dcyl}

The canonical quantization of the Dirac equation in cylindrical
coordinates has been studied in detail by Balantekin and
DeWeerd~\cite{baha} using the eigenstates of the transverse helicity
operators to obtain a complete set of solutions.  Here, for
uniformity of presentation we will use the tetrad formalism to
obtain a related decomposition in terms of Bessel functions of 1/2
integer order. Consider the change of variables $(\xi^\alpha
\rightarrow x^a)$ :
\begin{align}
   t = & \ t \>, \qquad z = z \>,
   \\ \notag
   x = & \ \rho \, \cos \theta
   \>,
   \qquad
   y = \rho \, \sin \theta
   \>.
\end{align}
with the metric $g_{ab} = \mathrm{diag}(-1,1,\rho^2,1)$, and
$\sqrt{-g}=\rho$.


\subsection{Cartesian tetrad representation}

The Vierbeins in \emph{fixed}-tetrad representation are calculated
as
\begin{align*}
   \vier{\alpha}{a}
   \ = \
   \left [
   \begin{array}{cccc}
   1 & 0 & 0 & 0 \\
   0 & \cos \theta & - \rho \, \sin \theta & 0 \\
   0 & \sin \theta &   \rho \, \cos \theta & 0 \\
   0 & 0 & 0 & 1
   \end{array}
   \right ]
   \>,
\end{align*}
whereas the inverse Vierbeins are obtained as
\begin{align*}
   \vinv{\alpha}{a}
   \ = \
   \left [
   \begin{array}{cccc}
   1 & 0 & 0 & 0 \\
   0 & \cos \theta & \sin \theta & 0 \\
   0 & - \displaystyle{\frac{1}{\rho}} \, \sin \theta &
         \displaystyle{\frac{1}{\rho}} \, \cos \theta & 0 \\
   0 & 0 & 0 & 1
   \end{array}
   \right ]
   \>.
\end{align*}
The Dirac equation becomes
\begin{equation}
   \bigl( \tilde \gamma^a \, \partial_a
   \, + \,
   M \bigr ) \ \psi
   \ = \ 0
\label{dirac_cyl}
   \>,
\end{equation}
with the gamma matrices, $\tilde \gamma^a = \vinv{\alpha}{a}
\gamma^\alpha$, defined as
\begin{align}
   \tilde \gamma^0 = & \, \gamma^0
   \>,
   \quad
   \tilde \gamma^3 = \gamma^3
   \>,
   \\
   \tilde \gamma^1 = & \
   \gamma^1 \cos\theta \, + \, \gamma^2 \sin \theta
   \>,
   \\
   \tilde \gamma^2 = & \
   \frac{1}{\rho} \
   \bigl ( - \gamma^1 \sin\theta \, + \, \gamma^2 \cos \theta
   \bigr )
   \>.
\end{align}

Following Villalba~\cite{villalba}, we write Eq.~\eqref{dirac_cyl}
as
\begin{align}
   \bigl ( \mathcal{K}_1 \ + \ \mathcal{K}_2 \bigr ) \,
   \Phi \ = \ 0
   \>,
   \quad \Phi = \mathrm{i}\, \gamma^1 \gamma^2 \, \psi
\label{dirac_eq1}
\end{align}
where $\mathcal{K}_1$ and $\mathcal{K}_2$ are the first-order
differential operators
\begin{align}
   \mathcal{K}_1(t,z) = & \
   \bigl (
      \tilde \gamma^0 \, \partial_t
      +
      \tilde \gamma^3 \, \partial_z
   \bigr ) \, \mathrm{i}\, \gamma^1 \gamma^2
\label{kcal1}
   \>,
   \\
   \mathcal{K}_2(\rho,\theta) = & \
   \bigl (
      \tilde \gamma^1 \, \partial_\rho
      +
      \tilde \gamma^2 \, \partial_\theta
      + M
   \bigr ) \, \mathrm{i}\, \gamma^1 \gamma^2
\label{kcal2}
   \>,
\end{align}
and
\begin{align}
   \mathrm{i}\, \gamma^1 \gamma^2
   =
   -
   \left [
   \begin{array}{cc}
   \sigma_3 & 0 \\
   0 & \sigma_3
   \end{array}
   \right ]
   .
\end{align}
Since the operators $\mathcal{K}_1$ and $\mathcal{K}_2$
commute~\cite{Kcomment}, $[\mathcal{K}_1,\, \mathcal{K}_2]$=0, we
can write the eigenvalue equations
\begin{align}
   \mathcal{K}_1 \, \Phi \ = \
   - \, \mathcal{K}_2 \, \Phi \ = \
   - \lambda \, \Phi
\label{eq:eigen}
   \>.
\end{align}
We will find the solution of Eq.~\eqref{dirac_eq1} has the form
\begin{align}
   \Phi(t,\rho,\theta,z) =
   \int \ k_\perp \, \mathrm{d}k_\perp \
   \Phi^{k_\perp}(t,\rho,\theta,z)
\label{phi_perp}
   \>,
\end{align}
where $\lambda^2 - M^2 = k_\perp^2$.


\paragraph{$\mathcal{K}_2$ sector.} The structure of the amplitudes
$\Phi^{k_\perp}$ is established by solving the eigenvector problem
for the operator $\mathcal{K}_2$. We have
\begin{equation}
   \bigl (
   \mathcal{K}_2
   \, - \,
   \lambda
   \bigr ) \ \Phi^{k_\perp}(\rho,\theta)
   \ = \ 0
\label{eq:K3caleq}
   \>,
\end{equation}
or
\begin{align}
   \bigl (
   \mathrm{i}\, \gamma^2 \, \partial_x
   - \mathrm{i}\, \gamma^1 \, \partial_y
   +  \mathrm{i}\, \gamma^1 \gamma^2 \, M
   - \lambda
   \bigr ) \,
   \Phi^{k_\perp}
   \ = \ 0
   \>,
\end{align}
which gives
\begin{align}
   \left [
   \begin{array}{cccc}
   - \lambda_+    & 0             & 0               & \mathcal{P}_- \\
   0              & - \lambda_-   & - \mathcal{P}_+ & 0 \\
   0              & \mathcal{P}_- & - \lambda_+     & 0 \\
   -\mathcal{P}_+ & 0             & 0               & - \lambda_-
   \end{array}
   \right ]
   \left [
   \begin{array}{c}
      \Phi^{k_\perp;1}(\rho,\theta) \\ \Phi^{k_\perp;2}(\rho,\theta) \\
      \Phi^{k_\perp;3}(\rho,\theta) \\ \Phi^{k_\perp;4}(\rho,\theta)
   \end{array}
   \right ]
   = 0 \>,
\label{eqrho}
\end{align}
where we have introduced the notations $\lambda_\pm = \lambda \pm
M$. The raising and lowering operators are defined as
\begin{align}
   \mathcal{P}_- = & \
   \partial_x - \mathrm{i} \partial_y
   =
   e^{-\mathrm{i}\theta} \,
   \Bigl (
      \partial_\rho - \frac{\mathrm{i}}{\rho} \, \partial_\theta
   \Bigr )
   \>,
   \\
   \mathcal{P}_+ = & \
   \partial_x + \mathrm{i} \partial_y
   =
   e^{\mathrm{i}\theta} \,
   \Bigl (
      \partial_\rho + \frac{\mathrm{i}}{\rho} \, \partial_\theta
   \Bigr )
   \>,
\end{align}
with the properties
\begin{align}
   \mathcal{P}_- \,
   \Bigl [
      e^{\mathrm{i}m\theta} \, J_{m}(k_\perp \rho)
   \Bigr ]
   = & \ + k_\perp \,
   \Bigl [
      e^{\mathrm{i}(m-1)\theta} \, J_{m-1}(k_\perp \rho)
   \Bigr ]
   \>,
   \\
   \mathcal{P}_+ \,
   \Bigl [
      e^{\mathrm{i}m\theta} \, J_{m}(k_\perp \rho)
   \Bigr ]
   = & \ - k_\perp \,
   \Bigl [
      e^{\mathrm{i}(m+1)\theta} \, J_{m+1}(k_\perp \rho)
   \Bigr ]
   \>.
\end{align}
The system of equations~\eqref{eqrho} decouple, and we obtain
\begin{align}
   \bigl [ \mathcal{P}_+ \mathcal{P}_- \ + \
   \lambda_+\lambda_-
   \bigr ] \
   \Phi^{k_\perp;2[4]}(\rho,\theta)
\label{eq_24}
   = & \ 0
   \>,
   \\
   \bigl [ \mathcal{P}_- \mathcal{P}_+ \ + \
   \lambda_+\lambda_-
   \bigr ] \
   \Phi^{k_\perp;3[1]}(\rho,\theta)
\label{eq_13}
   = & \ 0
   \>.
\end{align}
The spinor $\Phi^{k_\perp}(\rho,\theta)$ has the form
\begin{align}
   \Phi^{k_\perp}(\rho,\theta)
   =
   \red{
   \sum_{m=-\infty}^\infty \
   }
   \Phi^{k_\perp}_m(\rho,\theta)
   \>,
\end{align}
with
\begin{align}
   \Phi^{k_\perp}_m(\rho,\theta)
   =
   e^{\mathrm{i}m\theta} \,
   \left [
   \begin{array}{r}
      A^{k_\perp}\, e^{-\mathrm{i}\theta/2} \, J_{m-\frac{1}{2}}(k_\perp\rho) \\
      B^{k_\perp}\, e^{\mathrm{i}\theta/2} \, J_{m+\frac{1}{2}}(k_\perp\rho) \\
      C^{k_\perp}\, e^{-\mathrm{i}\theta/2} \, J_{m-\frac{1}{2}}(k_\perp\rho) \\
      D^{k_\perp}\, e^{\mathrm{i}\theta/2} \, J_{m+\frac{1}{2}}(k_\perp\rho)
   \end{array}
   \right ]
   \>,
\end{align}
and obtain the constraints
\begin{align}
   D^{k_\perp} = & \
   \frac{\lambda_+}{k_\perp} \, A^{k_\perp}
   =
   \frac{k_\perp}{\lambda_-} \, A^{k_\perp}
   \>,
   \\
   B^{k_\perp} = & \
   \frac{k_\perp}{\lambda_-} \, C^{k_\perp}
   =
   \frac{\lambda_+}{k_\perp} \, C^{k_\perp}
   \>.
\end{align}
We also find
\begin{equation}
   \lambda_+\lambda_- =
   \lambda^2 - M^2 = k_\perp^2
   \>,
\label{assumpt}
\end{equation}
which gives
\begin{equation}
   \lambda^2  = \omega_\perp^2 = M^2 + k_\perp^2
   \>,
   \>\>
   \textrm{or} \>\>
   \lambda = \pm \, \omega_\perp
   \>.
\end{equation}
Corresponding to the two eigenvalues, $\lambda=\pm \, \omega_\perp$,
we have two linearly-independent eigenvectors
\begin{align}
   \Phi^{k_\perp;s}_m(\rho,\theta)
   =
   e^{\mathrm{i}m\theta} \,
   \left [
   \begin{array}{r}
      A^{k_\perp}\, e^{-\mathrm{i}\theta/2} \, J_{m-\frac{1}{2}}(k_\perp\rho) \\
      \frac{k_\perp}{\lambda_-} \, C^{k_\perp}\,
      e^{\mathrm{i}\theta/2} \, J_{m+\frac{1}{2}}(k_\perp\rho) \\
      C^{k_\perp}\, e^{-\mathrm{i}\theta/2} \, J_{m-\frac{1}{2}}(k_\perp\rho) \\
      \frac{k_\perp}{\lambda_-} \, A^{k_\perp}\,
      e^{\mathrm{i}\theta/2} \, J_{m+\frac{1}{2}}(k_\perp\rho)
   \end{array}
   \right ]
\label{k1_eigv}
   ,
\end{align}
with $\lambda = s \, \omega_\perp$, and $s=\pm 1$.


\paragraph{$\mathcal{K}_1$ sector.} The amplitudes $A^{k_\perp}$
and $C^{k_\perp}$ are determined by solving the eigenvalue equation
for the operator~$\mathcal{K}_1$, i.e.
\begin{equation}
   \bigl (
   \mathcal{K}_1
   \, + \,
   \lambda
   \bigr ) \ \Phi^{k_\perp;s}_m(t,z)
   \ = \ 0
   \>,
\end{equation}
or
\begin{align}
   \Bigl [ \,
   \mathrm{i}\, \gamma^0 \gamma^1 \gamma^2 \, \partial_t
   + \mathrm{i}\, \gamma^3 \gamma^1 \gamma^2 \, \partial_z
   + \lambda \,
   \Bigr ] \,
   \Phi^{k_\perp;s}_m(t,z)
   \ = \ 0
   \>.
\end{align}
Once again using the representation~\eqref{gammas} of the Dirac
gamma matrices, we obtain the anticommuting products of gamma
matrices:
\begin{align}
   \mathrm{i}\, \gamma^0 \gamma^1 \gamma^2
   =
   \left [
   \begin{array}{cc}
   0 & - \sigma_3 \\
   \sigma_3 & 0
   \end{array}
   \right ]
   ,
   \quad
   \mathrm{i}\, \gamma^3 \gamma^1 \gamma^2
   =
   \left [
   \begin{array}{cc}
   0 & -1 \\
   -1 & 0
   \end{array}
   \right ]
   .
\end{align}
Since $\partial_t$ and $\partial_z$ commute with the above equation,
we can also Fourier transform the Dirac field in $t$ and $z$. We
introduce the field as
\begin{align}
   \Phi^{k_\perp;s}_m
   =
   \int \red{\mathrm{d}E} \ e^{\mathrm{i}E t}
   \int \red{\mathrm{d}k_z} \ e^{-\mathrm{i}k_z z} \
   \Phi^{k_\perp;s}_{m,k_z;E}
   \>,
\end{align}
which gives
\begin{align}
   \biggl \{ \!
   \left [
   \begin{array}{cc}
   0 & - \sigma_3 \!\! \\
   \!\! \sigma_3 & 0
   \end{array}
   \right ] \!\!
   ( \mathrm{i}E )
   & \! + \!
   \left [
   \begin{array}{cc}
   0 & -1 \!\! \\
   \!\! -1 & 0
   \end{array}
   \right ] \!
   ( -\mathrm{i}k_z )
   \! + \!
   \lambda
   \biggr \}
   \Phi^{k_\perp}_{m,k_z;E}
   = 0
\label{mk_eigzt}
   \>.
\end{align}
Eq.~\eqref{mk_eigzt} is equivalent to
\begin{align}
   \left [
   \begin{array}{cc}
   \lambda & \mathrm{i}(k_z-E) \\
   \mathrm{i}(k_z+E) & \lambda
   \end{array}
   \right ]
   \left [
   \begin{array}{r}
      A^{k_\perp}_{k_z,E} \\
      C^{k_\perp}_{k_z,E}
   \end{array}
   \right ]
   = 0
   \>.
\end{align}
For a nontrivial solution, we require the determinant to be zero,
which gives the usual dispersion relations
\begin{equation}
   E =
   \pm \, \sqrt{ \omega_\perp^2 + k_z^2 }
   = \pm \, \sqrt{ M^2 + \mathbf{k}^2 }
   \>.
\label{disp}
\end{equation}
We obtain
\begin{align}
   C^{k_\perp}_{k_z,E}
   = \frac{\mathrm{i}\lambda}{k_z-E} \ A^{k_\perp}_{k_z,E}
   = \frac{-\mathrm{i}(k_z+E)}{\lambda} \ A^{k_\perp}_{k_z,E}
   \>.
\end{align}
Hence, the solution to the free-field Dirac equation can be written
as
\begin{align}
   &
   \psi^{k_\perp;s}_{m,k_z;E}(t,\rho,\theta,z)
   = \mathrm{i} \, \gamma^1 \gamma^2 \
     \Phi^{k_\perp;s}_{m,k_z;E}(t,\rho,\theta,z)
\label{cyl_cart}
   \\ \notag &
   = \,
   N^{k_\perp;s}_{k_z,E} \,
   e^{\mathrm{i}( m\theta -k_z z )}
   \left [
   \begin{array}{r}
      - e^{-\mathrm{i}\theta/2} \, J_{m-\frac{1}{2}}(k_\perp\rho) \\
      \frac{k_\perp}{\lambda - M} \, \frac{\mathrm{i}\lambda}{k_z-E} \
      e^{\mathrm{i}\theta/2} \, J_{m+\frac{1}{2}}(k_\perp\rho) \\
      - \frac{\mathrm{i}\lambda}{k_z-E} \
      e^{-\mathrm{i}\theta/2} \, J_{m-\frac{1}{2}}(k_\perp\rho) \\
      \frac{k_\perp}{\lambda - M} \
      e^{\mathrm{i}\theta/2} \, J_{m+\frac{1}{2}}(k_\perp\rho)
   \end{array}
   \right ]
   .
\end{align}

The canonical quantization of the Dirac field in cylindrical
coordinates is obtained as
\begin{align}
   \hat \psi(t,\rho,\theta,z)
   = &
   \int \red{k_\perp \, \mathrm{d}k_\perp}
   \int \red{\mathrm{d}k_z}
   \sum_{m=-\infty}^\infty \sum_s
   \\ \notag & \times
   \Bigl [
   \psi^{k_\perp;s\, (+)}_{m,k_z} \,
       a^{k_\perp;s}_{m,k_z} +
   \psi^{k_\perp;s\, (-)}_{m,k_z} \,
       b^{k_\perp;s\, \dag}_{m,k_z}
   \Bigr ]
   \>.
\end{align}
where $\psi^{k_\perp;s\, (\pm)}_{m,k_z}$ correspond to the positive
and negative energy solutions, respectively. The normalization
constants $N^{k_\perp;s}_{k_z,(\pm)|E|}$ are obtained by requiring
the fields $\psi^{k_\perp;s\, (\pm)}_{m,k_z}$ to satisfy the
orthonormality relations
\begin{align}
   \bigl ( \bar \psi^{k_\perp;s(\pm)}_{m,k_z},
                \psi^{k_\perp';s'(\pm)}_{m',k_z'}
   \bigr )
   & =
   \red{
   \frac{\delta(k_\perp'-k_\perp)}{k_\perp} \ \delta(k_z'-k_z) \,
   }
   \delta_{m' m} \delta_{s' s}
   ,
\label{plus_cyl}
   \\
   \bigl ( \bar \psi^{k_\perp;s(+)}_{m,k_z},
                \psi^{k_\perp';s'(-)}_{m',k_z'}
   \bigr )
   & = 0
   \>.
\end{align}
where we have introduced the inner product
\begin{align}
   \bigl ( \bar \psi^{k_\perp;s}_{m,k_z}, \, &
                \psi^{k_\perp';s'}_{m',k_z'}
   \bigr )
   = -
   \int \rho\, \mathrm{d}\rho
   \int \mathrm{d}\theta
   \int \mathrm{d}z \
   \bar \psi^{k_\perp;s}_{m,k_z} \, \tilde \gamma^0 \,
        \psi^{k_\perp';s'}_{m',k_z'}
   \notag \\ = &
\label{dotprod_cyl}
   \int \rho\, \mathrm{d}\rho
   \int \mathrm{d}\theta
   \int \mathrm{d}z \
   \psi^{k_\perp;s\dag}_{m,k_z} \
        \psi^{k_\perp';s'}_{m',k_z'}
   \>.
\end{align}
Using the identities
\begin{align}
   2\pi \ \delta_{m'm}
   = & \
   \int \mathrm{d}\theta \ e^{\mathrm{i}(m'-m)\,\theta}
   \>,
   \\
   2\pi \ \delta(k_z' - k_z)
   = & \
   \int \mathrm{d}z \ e^{\mathrm{i}(k_z'-k_z)\,z}
   \>,
   \\
   \frac{\delta(k_\perp' - k_\perp)}{k_\perp}
   = & \
   \int \rho\, \mathrm{d}\rho \
   J_\nu(k_\perp' \rho) \, J_\nu(k_\perp \rho)
   \>,
\end{align}
we obtain the normalization constant
\begin{equation}
   N^{k_\perp;s}_{k_z, E} =
   \red{\frac{1}{2\pi}} \
   \sqrt{ \frac{E-k_z}{2 E} } \
   \sqrt{ \frac{\lambda-M}{2 \lambda} }
\label{norm_cyl}
   \>,
\end{equation}
for either positive or negative values of $E$ and $\lambda$. This
completes our derivation.


\subsection{Diagonal tetrad representation}

In cylindrical coordinates, we introduce the similarity
transformation
\begin{equation}
   S_\rho
          \ = \ \cos \frac{\theta}{2}
                \ - \ \gamma^1 \gamma^2 \, \sin \frac{\theta}{2}
          \ = \ \exp \Bigl ( - \frac{\theta}{2} \ \gamma^1 \gamma^2
                     \Bigr )
   \>,
\end{equation}
such that
\begin{align}
   S_\rho^{-1} \ \tilde \gamma^a \ S_\rho \ = \ \bar \gamma^a
   \>,
\end{align}
where
\begin{align}
   \bar \gamma^0 = \gamma^0 \>,
   \quad
   \bar \gamma^1 = \gamma^1 \>,
   \quad
   \bar \gamma^2 = \frac {1}{\rho} \ \gamma^2 \>,
   \quad
   \bar \gamma^3 = \gamma^3 \>.
\end{align}
The diagonal Vierbein representation, $\bar \gamma^\alpha =
\vvinv{\alpha}{a} \gamma^\alpha$, is introduced as
\begin{align}
   \vvier{\alpha}{a} \ = \ & \mathrm{diag}(1,1,\rho,1)
   \>,
   \\
   \vvinv{\alpha}{a} \ = \ &
   \mathrm{diag}\bigl ( 1,1,\frac{1}{\rho},1 \bigr)
   \>,
\end{align}
and the Dirac equation can be written as
\begin{equation}
   \bigl( \bar \gamma^a \, \nabla_a
   \, + \,
   M \bigr ) \ \psi_{\mathrm{[d]}}
   \ = \ 0
\label{dirac_cylb}
   \>,
\end{equation}
where we have introduced the diagonal Dirac wave function
\begin{equation}
   \psi_{\mathrm{[d]}} \ = \
   S_\rho^{-1} \ \psi
\label{srho_psi}
   \>.
\end{equation}
In the cylindrical system of coordinates, the covariant derivatives
\begin{equation}
   \nabla_a \ = \
   S_\rho^{-1} \ \partial_a \ S_\rho
   \ = \
   \partial_a \ - \
   \Gamma_a
   \>,
\end{equation}
Here, the \emph{spinor} connection $\Gamma_a$,
\begin{align}
   \Gamma_\theta = & \ \frac{1}{2} \, \gamma^1 \gamma^2
   \>,
   \quad
   \Gamma_t = \Gamma_\rho = \Gamma_z =
   0
   \>,
\end{align}
are obtained using the (nonzero) Christoffel symbols
\begin{align}
   \Gamma^\rho_{\theta \theta} = & \ - \rho \>,
   \\
   \Gamma^\theta_{\rho \theta} = \Gamma^\theta_{\theta \rho} = & \
   \frac{1}{\rho}
   \>.
\end{align}
The covariant derivatives are obtained as
\begin{align}
   \nabla_t = \partial_t \>,
   \quad
   \nabla_\rho = \partial_\rho \>,
   \\ \notag
   \nabla_\theta = \partial_\theta - \frac{1}{2} \, \gamma^1 \gamma^2
   \>,
   \quad
   \nabla_z = \partial_z
   \>.
\end{align}
With the above definitions, the Dirac equation~\eqref{dirac_cylb}
can be written as
\begin{align}
   \Bigl [
     \gamma^0 \, \partial_t
   + \gamma^1 \, \Bigl ( \partial_\rho + \frac{1}{2\rho} \Bigr )
   + \gamma^2 \, \frac{1}{\rho} \ \partial_\theta
   + \gamma^3 \, \partial_z
   + M
   \Bigr ] \, \psi_{\mathrm{[d]}} = 0
   \>.
\label{dyrac_cylb1}
\end{align}
Following the same approach to the separation of variables as in the
case of the Cartesian tetrad representation, with the difference
that now the operator $\mathcal{K}_2$ is
\begin{align}
   \mathcal{K}_2(\rho,\theta) = & \
   \Bigl [
   \gamma^1 \, \Bigl ( \partial_\rho + \frac{1}{2\rho} \Bigr )
   + \gamma^2 \, \frac{1}{\rho} \ \partial_\theta
   + M
   \Bigr ] \, \mathrm{i}\, \gamma^1 \gamma^2
\label{kcal2_diag}
   \>.
\end{align}
We find that the solution of Eq.~\eqref{dyrac_cylb1} has the form
\begin{align}
   \psi_{E\mathrm[d]} =
   \int \red{k_\perp \, \mathrm{d}k_\perp}
   \int \red{\mathrm{d}k_z}
   \sum_{m=-\infty}^\infty \sum_s \
   \psi^{k_\perp;s}_{m,k_z;E\mathrm[d]}
   \>,
\end{align}
where the function~$\psi^{k_\perp;s\mathrm[d]}_{m,k_z;E}$ is given
by
\begin{align}
   &
   \psi^{k_\perp;s}_{m,k_z;E\mathrm[d]}(t,\rho,\theta,z)
   = \mathrm{i} \, \gamma^1 \gamma^2 \
     \Phi^{k_\perp;s}_{m,k_z;E\mathrm[d]}(t,\rho,\theta,z)
\label{cyl_diag}
   \\ \notag &
   = \,
   N^{k_\perp;s}_{k_z,E} \,
   e^{\mathrm{i}( m\theta -k_z z )}
   \left [
   \begin{array}{r}
      - J_{m-\frac{1}{2}}(k_\perp\rho) \\
      \frac{k_\perp}{\lambda - M} \, \frac{\mathrm{i}\lambda}{k_z-E} \
      J_{m+\frac{1}{2}}(k_\perp\rho) \\
      - \frac{\mathrm{i}\lambda}{k_z-E} \
      J_{m-\frac{1}{2}}(k_\perp\rho) \\
      \frac{k_\perp}{\lambda - M} \
      J_{m+\frac{1}{2}}(k_\perp\rho)
   \end{array}
   \right ]
   ,
\end{align}
where $N^{k_\perp;s}_{k_z,E}$ is given by Eq.~\eqref{norm_cyl}. The
Dirac fields in the Cartesian and diagonal tetrad representations
(see Eqs.~\eqref{cyl_cart} and~\eqref{cyl_diag}, respectively),
respectively are related via the similarity transformation
introduced in Eq.~\eqref{srho_psi}. As advertised, we obtain
\begin{align}
   S_\rho =
   \left [
   \begin{array}{cccc}
   e^{-\mathrm{i}\theta/2} & 0 & 0 & 0 \\
   0 & e^{\mathrm{i}\theta/2} & 0 & 0  \\
   0 & 0 & e^{-\mathrm{i}\theta/2} & 0\\
   0 & 0 & 0 & e^{\mathrm{i}\theta/2}\\
   \end{array}
   \right ]
\label{eq:srho}
   ,
\end{align}
which leads to
\begin{equation}
   \psi^{k_\perp;s}_{m,k_z;E} = \
   S_\rho \ \psi^{k_\perp;s}_{m,k_z;E\mathrm[d]}
   \>.
\end{equation}


\section{Boost-invariant coordinates with azimuthal symmetry}
\label{sec:4d}

Consider the change of variables $(\xi^\alpha \rightarrow x^a)$ :
\begin{align}
   t = & \ \tau \, \cosh \eta
   \>,
   \quad
   z = \tau \, \sinh \eta
   \>,
   \\ \notag
   x = & \ \rho \, \cos \theta
   \>,
   \quad
   y = \rho \, \sin \theta
   \>,
\end{align}
which gives the metric $g_{ab} = \mathrm{diag}(-1,1,\rho^2,\tau^2)$,
and $\sqrt{-g}=\tau \rho$.

Since the $(x,y)$ and $(t,z)$ subspaces do not interact, the
transformations we need are indeed very simple. The key ingredients
are:
\begin{itemize}
   \item Cartesian Vierbeins:
\begin{align*}
   \vier{\alpha}{a}
   \ = \
   \left [
   \begin{array}{cccc}
   \cosh \eta & 0 & 0 & \tau \, \sinh \eta \\
   0 & \cos \theta & - \rho \, \sin \theta & 0 \\
   0 & \sin \theta &   \rho \, \cos \theta & 0 \\
   \sinh \eta & 0 & 0 &   \tau \, \cosh \eta
   \end{array}
   \right ]
   \>,
\end{align*}
and
\begin{align*}
   \vinv{\alpha}{a}
   \ = \
   \left [
   \begin{array}{cccc}
   \cosh \eta & 0 & 0 & - \sinh \eta \\
   0 & \cos \theta & \sin \theta & 0 \\
   0 & - \displaystyle{\frac{1}{\rho}} \, \sin \theta &
         \displaystyle{\frac{1}{\rho}} \, \cos \theta & 0 \\
   - \displaystyle{\frac{1}{\tau}} \, \sinh \eta & 0 & 0 &
         \displaystyle{\frac{1}{\tau}} \, \cosh \eta
   \end{array}
   \right ]
   \>.
\end{align*}

   \item gamma matrices in the fixed-tetrad
   representation:
\begin{align}
   \tilde \gamma^0 = & \
   \gamma^0 \cosh\eta \, - \, \gamma^3 \sinh \eta
   \>,
   \\
   \tilde \gamma^3 = & \
   \frac{1}{\tau} \
   \bigl ( - \gamma^0 \sinh\eta \, + \, \gamma^3 \cosh \eta
   \bigr )
   \\
   \tilde \gamma^1 = & \
   \gamma^1 \cos\theta \, + \, \gamma^2 \sin \theta
   \>,
   \\
   \tilde \gamma^2 = & \
   \frac{1}{\rho} \
   \bigl ( - \gamma^1 \sin\theta \, + \, \gamma^2 \cos \theta
   \bigr )
   \>.
\end{align}

    \item similarity transformation: $S=S_\tau S_\rho$. Note that
    the individual similarity transformations commute with each other.

    \item diagonal Vierbeins:
\begin{align}
   \vvier{\alpha}{a} \ = \ & \mathrm{diag}(1,1,\rho,\tau)
   \>,
   \\
   \vvinv{\alpha}{a} \ = \ &
   \mathrm{diag}\bigl ( 1,1,\frac{1}{\rho},\frac{1}{\tau} \bigr)
   \>,
\end{align}

    \item gamma matrices in the rotating-tetrad
    representation: $\bar \gamma^a = S^{-1} \, \tilde \gamma^a \,
    S$, where
\begin{align}
   \bar \gamma^0 = \gamma^0 \>,
   \quad
   \bar \gamma^1 = \gamma^1 \>,
   \quad
   \bar \gamma^2 = \frac {1}{\rho} \, \gamma^2 \>,
   \quad
   \bar \gamma^3 = \frac {1}{\tau} \, \gamma^3 \>.
\end{align}

    \item diagonal Dirac field:
\begin{equation}
   \psi_{\mathrm{[d]}} \ = \
   S^{-1} \ \psi
   \ = \
   S_\tau^{-1} \,
   S_\rho^{-1} \ \psi
\label{eq:similar}
   \>.
\end{equation}

   \item covariant derivatives:
\begin{align}
   \nabla_\tau = \partial_\tau \>,
   \quad
   \nabla_\rho = \partial_\rho \>,
   \\ \notag
   \nabla_\theta = \partial_\theta - \frac{1}{2} \, \gamma^1 \gamma^2
   \>,
   \quad
   \nabla_\eta = \partial_\eta - \frac{1}{2} \, \gamma^0 \gamma^3
   \>.
\end{align}

\end{itemize}


\subsection{Cartesian tetrad representation}

The Dirac equation
\begin{equation}
   \bigl( \tilde \gamma^a \, \partial_a
   \, + \,
   M \bigr ) \ \psi
   \ = \ 0
   \>,
\end{equation}
is solved using the separation of variables procedure outlined
before, see Eqs.~\eqref{dirac_eq1}, \eqref{kcal1} and~\eqref{kcal2}.
The modified Dirac field $\Phi(\tau,\rho,\theta,\eta)$ has the
form~\eqref{phi_perp}, and we solve the eigenvalue problems defined
in Eq.~\eqref{eq:eigen}. Again, since there is no interaction
between the $(x,y)$ and $(t,z)$ subspaces, it follows that the
$(\rho,\theta)$ structure of $\Phi(\tau,\rho,\theta,\eta)$ remains
the same as in~\eqref{k1_eigv}, and we only need to concern
ourselves only with solving the eigenvalue problem for the operator
$\mathcal{K}_1$ in the new set of coordinates, $(\tau,\eta)$.
Explicitly, we seek the amplitudes $A^{k_\perp}_{k_\eta}(\tau)$ and
$C^{k_\perp}_{k_\eta}(\tau)$ required by~\eqref{k1_eigv}.

Consider the eigenvalue problem
\begin{equation}
   \bigl (
   \mathcal{K}_1
   \, + \,
   \lambda
   \bigr ) \ \Phi^{k_\perp}(\tau)
   \ = \ 0
   \>.
\end{equation}
Using the raising and lowering operators defined in~\eqref{tplus}
and~\eqref{tminus}, respectively, the above equation reads
\begin{align}
   \left [
   \begin{array}{cccc}
   \lambda & 0 &  \!\! - \mathcal{T}_- & 0 \\
   0 & \lambda & 0 & \mathcal{T}_+ \\
   \mathcal{T}_+& 0 & \lambda & 0 \\
   0 &  \!\! - \mathcal{T}_- & 0 & \lambda
   \end{array}
   \right ] \!\!
   \left [
   \begin{array}{r}
      A^{k_\perp}_{k_\eta}(\tau)
      e^{-\mathrm{i}\theta/2} J_{m-\frac{1}{2}}(k_\perp\rho) \\
      \frac{k_\perp}{\lambda_-} C^{k_\perp}_{k_\eta}(\tau)
      e^{\mathrm{i}\theta/2} J_{m+\frac{1}{2}}(k_\perp\rho) \\
      C^{k_\perp}_{k_\eta}(\tau)
      e^{-\mathrm{i}\theta/2} J_{m-\frac{1}{2}}(k_\perp\rho) \\
      \frac{k_\perp}{\lambda_-} A^{k_\perp}_{k_\eta}(\tau)
      e^{\mathrm{i}\theta/2} J_{m+\frac{1}{2}}(k_\perp\rho)
   \end{array}
   \right ] \!\!
   = 0 ,
\label{mk_eig}
\end{align}
which is equivalent to
\begin{align}
   \left [
   \begin{array}{cc}
   \lambda & - \mathcal{T}_- \\
   \mathcal{T}_+ & \lambda
   \end{array}
   \right ]
   \left [
   \begin{array}{r}
      A^{k_\perp}_{k_\eta}(\tau) \\
      C^{k_\perp}_{k_\eta}(\tau)
   \end{array}
   \right ]
   = 0
   \>.
\end{align}
The solutions of the above equations are Bessel functions of
\emph{complex} order (and \emph{real}
argument)~\cite{bessel_complex}. So, similarly to
Eq.~\eqref{eq:psi_2dc}, the Dirac field in the Cartesian tetrad
representation becomes
\begin{align}
   \psi^{k_\perp;s}_{m,k_\eta}
   = \mathrm{i} & \, \gamma^1 \gamma^2 \
     \Phi^{k_\perp;s}_{m,k_\eta}
   =
   N^{k_\perp;s} \,
   e^{\mathrm{i}( m\theta + k_\eta \eta )}
   \\ \notag
   & \times
   \left [
   \begin{array}{r}
      - e^{-\mathrm{i}(\theta-\mathrm{i}\eta)/2} \,
      J_{\, \mathrm{i}k_\eta-\frac{1}{2}}(\lambda \tau)\, J_{m-\frac{1}{2}}(k_\perp\rho) \\
      \frac{k_\perp}{\lambda_-} \,
      e^{\mathrm{i}(\theta-\mathrm{i}\eta)/2} \,
      J_{\, \mathrm{i}k_\eta+\frac{1}{2}}(\lambda \tau)\, J_{m+\frac{1}{2}}(k_\perp\rho) \\
      - e^{-\mathrm{i}(\theta+\mathrm{i}\eta)/2} \,
      J_{\, \mathrm{i}k_\eta+\frac{1}{2}}(\lambda \tau)\, J_{m-\frac{1}{2}}(k_\perp\rho) \\
      \frac{k_\perp}{\lambda_-} \,
      e^{\mathrm{i}(\theta+\mathrm{i}\eta)/2} \,
      J_{\, \mathrm{i}k_\eta-\frac{1}{2}}(\lambda \tau)\, J_{m+\frac{1}{2}}(k_\perp\rho)
   \end{array}
   \right ]
   .
\end{align}
Requiring the usual positive/negative-energy boundary
conditions~\cite{disessa_bc}, we obtain
\begin{align}
   \psi^{k_\perp;s(+)}_{m,k_\eta}
   & =
   N^{k_\perp;s}_{(+)} \,
   e^{\mathrm{i}( m\theta + k_\eta \eta )}
   \\ \notag
   & \times
   \left [
   \begin{array}{r}
      - e^{-\mathrm{i}(\theta-\mathrm{i}\eta)/2} \,
      H^{(2)}_{\, \mathrm{i}k_\eta-\frac{1}{2}}(\lambda \tau)\, J_{m-\frac{1}{2}}(k_\perp\rho) \\
      \frac{k_\perp}{\lambda_-} \,
      e^{\mathrm{i}(\theta-\mathrm{i}\eta)/2} \,
      H^{(2)}_{\, \mathrm{i}k_\eta+\frac{1}{2}}(\lambda \tau)\, J_{m+\frac{1}{2}}(k_\perp\rho) \\
      - e^{-\mathrm{i}(\theta+\mathrm{i}\eta)/2} \,
      H^{(2)}_{\, \mathrm{i}k_\eta+\frac{1}{2}}(\lambda \tau)\, J_{m-\frac{1}{2}}(k_\perp\rho) \\
      \frac{k_\perp}{\lambda_-} \,
      e^{\mathrm{i}(\theta+\mathrm{i}\eta)/2} \,
      H^{(2)}_{\, \mathrm{i}k_\eta-\frac{1}{2}}(\lambda \tau)\, J_{m+\frac{1}{2}}(k_\perp\rho)
   \end{array}
   \right ]
   ,
\end{align}
and
\begin{align}
   \psi^{k_\perp;s(-)}_{m,k_\eta}
   & =
   N^{k_\perp;s}_{(-)} \,
   e^{\mathrm{i}( m\theta + k_\eta \eta )}
   \\ \notag
   & \times
   \left [
   \begin{array}{r}
      - e^{-\mathrm{i}(\theta-\mathrm{i}\eta)/2} \,
      H^{(1)}_{\, \mathrm{i}k_\eta-\frac{1}{2}}(\lambda \tau)\, J_{m-\frac{1}{2}}(k_\perp\rho) \\
      \frac{k_\perp}{\lambda_-} \,
      e^{\mathrm{i}(\theta-\mathrm{i}\eta)/2} \,
      H^{(1)}_{\, \mathrm{i}k_\eta+\frac{1}{2}}(\lambda \tau)\, J_{m+\frac{1}{2}}(k_\perp\rho) \\
      - e^{-\mathrm{i}(\theta+\mathrm{i}\eta)/2} \,
      H^{(1)}_{\, \mathrm{i}k_\eta+\frac{1}{2}}(\lambda \tau)\, J_{m-\frac{1}{2}}(k_\perp\rho) \\
      \frac{k_\perp}{\lambda_-} \,
      e^{\mathrm{i}(\theta+\mathrm{i}\eta)/2} \,
      H^{(1)}_{\, \mathrm{i}k_\eta-\frac{1}{2}}(\lambda \tau)\, J_{m+\frac{1}{2}}(k_\perp\rho)
   \end{array}
   \right ]
   .
\end{align}

The canonical quantization of the Dirac field is obtained as
\begin{align}
   \hat \psi(\tau,\rho,\theta,\eta)
   = &
   \int \red{k_\perp \, \mathrm{d}k_\perp}
   \int \red{\mathrm{d}k_\eta}
   \sum_{m=-\infty}^\infty \sum_s
   \\ \notag & \times
   \Bigl [
   \psi^{k_\perp;s\, (+)}_{m,k_\eta} \,
       a^{k_\perp;s}_{m,k_\eta} +
   \psi^{k_\perp;s\, (-)}_{m,k_\eta} \,
       b^{k_\perp;s\, \dag}_{m,k_\eta}
   \Bigr ]
   \>.
\end{align}
where $\psi^{k_\perp;s\, (\pm)}_{m,k_\eta}$ correspond to the
positive and negative energy solutions, respectively. The
normalization constants $N^{k_\perp;s}_{(\pm)}$ are obtained by
requiring the fields $\psi^{k_\perp;s\, (\pm)}_{m,k_\eta}$ to
satisfy the orthonormality relations
\begin{align}
   \bigl ( \bar \psi^{k_\perp;s(\pm)}_{m,k_\eta},
                \psi^{k_\perp';s'(\pm)}_{m',k_\eta'}
   \bigr )
   & =
   \red{
   \frac{\delta(k_\perp'-k_\perp)}{k_\perp} \ \delta(k_\eta'-k_\eta) \,
   }
   \delta_{m' m} \delta_{s' s}
   ,
\label{plus_bi}
   \\
   \bigl ( \bar \psi^{k_\perp;s(+)}_{m,k_\eta},
                \psi^{k_\perp';s'(-)}_{m',k_\eta'}
   \bigr )
   & = 0
   \>.
\end{align}
where we introduce the inner product
\begin{align}
   \bigl ( \bar \psi^{k_\perp;s}_{m,k_\eta}, \, &
                \psi^{k_\perp';s'}_{m',k_\eta'}
   \bigr )
   = - \tau
   \int \! \rho\, \mathrm{d}\rho \!
   \int \! \mathrm{d}\theta \!
   \int \! \mathrm{d}z \
   \bar \psi^{k_\perp;s}_{m,k_\eta} \, \tilde \gamma^0 \,
        \psi^{k_\perp';s'}_{m',k_\eta'}
\label{dotprod_bi}
   \>.
\end{align}
We find the normalization constants
\begin{equation}
   N^{k_\perp;s}_{(\pm)} =
   \red{\frac{1}{2\pi}} \
   \sqrt{ \frac{M\, (\lambda-M)}{16 \lambda} } \
   e^{ \pm \, \pi k_\eta/2 }
\label{norm_bi}
   \>.
\end{equation}


\subsection{Diagonal tetrad representation}

In the diagonal tetrad representation the Dirac equation takes the
form
\begin{align}
   \Bigl [
     \gamma^0 \, \Bigl ( \partial_\tau + \frac{1}{2\tau} & \Bigr )
   + \gamma^1 \, \Bigl ( \partial_\rho + \frac{1}{2\rho} \Bigr )
\label{dirac_eq}
   \\ \notag &
   + \gamma^2 \, \frac{1}{\rho} \ \partial_\theta
   + \gamma^3 \, \frac{1}{\tau} \ \partial_\eta
   + M
   \Bigr ] \, \psi_{\mathrm{[d]}} = 0
   \>,
\end{align}
and has a solution of the form
\begin{align}
   \psi_{\mathrm[d]} =
   \int \red{k_\perp \, \mathrm{d}k_\perp}
   \int \red{\mathrm{d}k_\eta}
   \sum_{m=-\infty}^\infty \sum_s \
   \psi^{k_\perp;s}_{m,k_\eta\mathrm{[d]}}
   \>,
\end{align}
with
\begin{align}
   \psi^{k_\perp;s}_{m,k_\eta\mathrm{[d]}}
   = \mathrm{i} & \, \gamma^1 \gamma^2 \
     \Phi^{k_\perp;s}_{m,k_\eta}
   =
   N^{k_\perp;s} \,
   e^{\mathrm{i}( m\theta + k_\eta \eta )}
   \\ \notag
   & \times
   \left [
   \begin{array}{r}
      -
      J_{\, \mathrm{i}k_\eta-\frac{1}{2}}(\lambda \tau)\, J_{m-\frac{1}{2}}(k_\perp\rho) \\
      \frac{k_\perp}{\lambda_-} \,
      J_{\, \mathrm{i}k_\eta+\frac{1}{2}}(\lambda \tau)\, J_{m+\frac{1}{2}}(k_\perp\rho) \\
      -
      J_{\, \mathrm{i}k_\eta+\frac{1}{2}}(\lambda \tau)\, J_{m-\frac{1}{2}}(k_\perp\rho) \\
      \frac{k_\perp}{\lambda_-} \,
      J_{\, \mathrm{i}k_\eta-\frac{1}{2}}(\lambda \tau)\, J_{m+\frac{1}{2}}(k_\perp\rho)
   \end{array}
   \right ]
   .
\end{align}
Then, the positive- and negative-energy solutions correspond to
\begin{align}
   \psi^{k_\perp;s(+)}_{m,k_\eta\mathrm{[d]}}
   = \ &
   N^{k_\perp;s}_{(+)} \,
   e^{\mathrm{i}( m\theta + k_\eta \eta )}
   \\ \notag
   & \times
   \left [
   \begin{array}{r}
      -
      H^{(2)}_{\, \mathrm{i}k_\eta-\frac{1}{2}}(\lambda \tau)\, J_{m-\frac{1}{2}}(k_\perp\rho) \\
      \frac{k_\perp}{\lambda_-} \,
      H^{(2)}_{\, \mathrm{i}k_\eta+\frac{1}{2}}(\lambda \tau)\, J_{m+\frac{1}{2}}(k_\perp\rho) \\
      -
      H^{(2)}_{\, \mathrm{i}k_\eta+\frac{1}{2}}(\lambda \tau)\, J_{m-\frac{1}{2}}(k_\perp\rho) \\
      \frac{k_\perp}{\lambda_-} \,
      H^{(2)}_{\, \mathrm{i}k_\eta-\frac{1}{2}}(\lambda \tau)\, J_{m+\frac{1}{2}}(k_\perp\rho)
   \end{array}
   \right ]
   ,
\end{align}
and
\begin{align}
   \psi^{k_\perp;s(-)}_{m,k_\eta\mathrm{[d]}}
   = \ &
   N^{k_\perp;s}_{(-)} \,
   e^{\mathrm{i}( m\theta + k_\eta \eta )}
   \\ \notag
   & \times
   \left [
   \begin{array}{r}
      -
      H^{(1)}_{\, \mathrm{i}k_\eta-\frac{1}{2}}(\lambda \tau)\, J_{m-\frac{1}{2}}(k_\perp\rho) \\
      \frac{k_\perp}{\lambda_-} \,
      H^{(1)}_{\, \mathrm{i}k_\eta+\frac{1}{2}}(\lambda \tau)\, J_{m+\frac{1}{2}}(k_\perp\rho) \\
      -
      H^{(1)}_{\, \mathrm{i}k_\eta+\frac{1}{2}}(\lambda \tau)\, J_{m-\frac{1}{2}}(k_\perp\rho) \\
      \frac{k_\perp}{\lambda_-} \,
      H^{(1)}_{\, \mathrm{i}k_\eta-\frac{1}{2}}(\lambda \tau)\, J_{m+\frac{1}{2}}(k_\perp\rho)
   \end{array}
   \right ]
   ,
\end{align}
where $N^{k_\perp;s}_{(\pm)}$ is given by Eq.~\eqref{norm_bi}. The
Dirac fields in the Cartesian and diagonal tetrad representations
are related via the similarity transformation~\eqref{eq:similar}. We
note that $S_\rho$ is given by~\eqref{eq:srho}, and $S_\tau$ is
defined as
\begin{align}
   S_\tau =
   \left [
   \begin{array}{cccc}
   e^{-\eta/2} & 0 & 0 & 0 \\
   0 & e^{\eta/2} & 0 & 0  \\
   0 & 0 & e^{\eta/2} & 0\\
   0 & 0 & 0 & e^{-\eta/2}\\
   \end{array}
   \right ]
\label{eq:stau}
   .
\end{align}
We can easily verify that indeed we have
\begin{equation}
   \psi^{k_\perp;s}_{m,k_\eta} = \
   S_\tau S_\rho \ \psi^{k_\perp;s}_{m,k_\eta\mathrm[d]}
   \>.
\end{equation}


\section{Conclusions}
\label{sec:concl}

In this paper we have shown step by step how to obtain the canonical
quantization in cylindrical geometry as well as in the 1+1
dimensional  fluid rapidity and proper time coordinates.  Assuming
that we are interested in a quantum back-reaction problem with a
background  chromoelectric field that is a function only of $\tau$,
then the separation of variables we have made will continue for the
problem of interest, and the initial value problem choice that
$A(\tau_0) = 0$, tells us that the correct mode expansion at
$\tau=\tau_0$ for the Dirac equation is given by the problem we have
solved.  Thus the canonical quantization presented here for the
Dirac equation, will allow us to determine the initial conditions
for the quark state (adiabatic vacuum) so that we will be able to
numerically solve for the production of quarks and the degradation
of the chromoelectric field in a manner similar to the problem
addressed by Cooper \emph{et al}. in Ref.~\cite{CEKMS}.  By
additionally having the correct modes for the transverse degrees of
freedom, we can exploit the cylindrical symmetry and reduce the
number of momentum modes needed in the 3+1 dimensional problems to
be $N^2$ instead of $N^3$. These numerical simulations will be soon
undertaken in the case of color SU(3) quarks.


\begin{acknowledgments}
The authors would like to thank A.B. Balantekin and A.J. DeWeerd for
making available their unpublished work, and we would also like to
thank the Santa Fe Institute for its hospitality during the
completion of this work.
\end{acknowledgments}


\appendix


\section{Comments on the solution of the Dirac equation in
cylindrical coordinates}

First, let us note that the same normalization constant~$N_{k_z,
E}^{k_\perp}$ is determined by requiring the conditions~\cite{baha}:
\begin{align}
   &
   \sum_{m=-\infty}^\infty \
   \Phi^{k_\perp;s\, \dag}_{m,k_z;|E|}(\rho,\theta) \
   \Phi^{k_\perp;s'}_{m,k_z;|E|}(\rho,\theta)
   = \delta_{ss'}
   \>,
   \\
   &
   \sum_{m=-\infty}^\infty \
   \Phi^{k_\perp;s\, \dag}_{m,k_z;-|E|}(\rho,\theta) \
   \Phi^{k_\perp;s'}_{m,k_z;-|E|}(\rho,\theta)
   = \delta_{ss'}
   \>.
\end{align}
The following conditions are enforced by construction
\begin{align}
   &
   \sum_{m=-\infty}^\infty \
   \Phi^{k_\perp;s\, \dag}_{m,k_z;|E|}(\rho,\theta) \
   \Phi^{k_\perp;s'}_{m,k_z;-|E|}(\rho,\theta)
   = 0
   \>,
   \\ &
   \sum_{m=-\infty}^\infty \
   \Phi^{k_\perp;s\, \dag}_{m,k_z;-|E|}(\rho,\theta) \
   \Phi^{k_\perp;s'}_{m,k_z;|E|}(\rho,\theta)
   = 0
   \>.
\end{align}
In using the above conditions one makes use of Bessel functions'
property~(see discussion regarding the Graf generalization of Neumann's formula in Ref.~\cite{watson}):
\begin{align}
   \sum_{m=-\infty}^{+\infty}
   J_{m + \nu}(x) \, J_{m}(y) \, e^{\pm \mathrm{i} m \theta}
   &=
   e^{\pm \mathrm{i} \nu \phi} \, J_{\nu}(z) \>,
\end{align}
where
\begin{align}
   z = \sqrt{ x^2 + y^2 - 2 x y \, \cos \theta } \>,
   \\
   x - y \, \cos \theta = z \, \cos \phi \>,
   \\
   y \, \sin \theta = z \, \sin \phi \>.
\end{align}

Secondly, we discuss the boundary conditions in the $(x,y)$
subspace. Provided that we require the solution to be regular at the
origin, then the generic form~\eqref{k1_eigv} reads as follows: for
$m\ge 1$, we have
\begin{align}
   \Phi^{k_\perp;s}_m(\rho,\theta)
   =
   e^{\mathrm{i}m\theta} \,
   \left [
   \begin{array}{r}
      A^{k_\perp}\, e^{-\mathrm{i}\theta/2} \, J_{m-\frac{1}{2}}(k_\perp\rho) \\
      \frac{k_\perp}{\lambda_-} \, C^{k_\perp}\,
      e^{\mathrm{i}\theta/2} \, J_{m+\frac{1}{2}}(k_\perp\rho) \\
      C^{k_\perp}\, e^{-\mathrm{i}\theta/2} \, J_{m-\frac{1}{2}}(k_\perp\rho) \\
      \frac{k_\perp}{\lambda_-} \, A^{k_\perp}\,
      e^{\mathrm{i}\theta/2} \, J_{m+\frac{1}{2}}(k_\perp\rho)
   \end{array}
   \right ]
   ,
\end{align}
whereas for $m\le -1$ we  have
\begin{align}
   \Phi^{k_\perp;s}_m(\rho,\theta)
   =
   e^{\mathrm{i}m\theta} \,
   \left [
   \begin{array}{r}
      A^{k_\perp}\, e^{-\mathrm{i}\theta/2} \, Y_{m-\frac{1}{2}}(k_\perp\rho) \\
      \frac{k_\perp}{\lambda_-} \, C^{k_\perp}\,
      e^{\mathrm{i}\theta/2} \, Y_{m+\frac{1}{2}}(k_\perp\rho) \\
      C^{k_\perp}\, e^{-\mathrm{i}\theta/2} \, Y_{m-\frac{1}{2}}(k_\perp\rho) \\
      \frac{k_\perp}{\lambda_-} \, A^{k_\perp}\,
      e^{\mathrm{i}\theta/2} \, Y_{m+\frac{1}{2}}(k_\perp\rho)
   \end{array}
   \right ]
   ,
\end{align}
where $Y_\nu(x)$ is the Neumann function. For integer~$m$ we have
the property~\cite{watson}
\begin{align}
   Y_{m-\frac{1}{2}}(x) = & \
   (-)^m \, J_{-m+\frac{1}{2}}(x)
   \>, \\
   Y_{m+\frac{1}{2}}(x) = & \
   (-)^{m+1} \, J_{-m-\frac{1}{2}}(x)
   \>.
\end{align}
Finally, when $m=0$ the solution~\eqref{k1_eigv} becomes
\begin{align}
   \Phi^{k_\perp;s}_m(\rho,\theta)
   =
   \left [
   \begin{array}{r}
      A^{k_\perp}\, e^{-\mathrm{i}\theta/2} \, J_{-\frac{1}{2}}(k_\perp\rho) \\
      \frac{k_\perp}{\lambda_-} \, C^{k_\perp}\,
      e^{\mathrm{i}\theta/2} \, Y_{\frac{1}{2}}(k_\perp\rho) \\
      C^{k_\perp}\, e^{-\mathrm{i}\theta/2} \, J_{-\frac{1}{2}}(k_\perp\rho) \\
      \frac{k_\perp}{\lambda_-} \, A^{k_\perp}\,
      e^{\mathrm{i}\theta/2} \, Y_{\frac{1}{2}}(k_\perp\rho)
   \end{array}
   \right ]
   .
\end{align}

It is interesting to note that the discussion of the
boundary conditions in the $(x,y)$ subspace simplifies dramatically
if we choose Bessel functions of integer rather than half-integer
order in~\eqref{k1_eigv}. In other words, in the $(x,y)$ subspace we
can replace the form~\eqref{k1_eigv} which involves Bessel functions
of half-integer order, and use instead the form
\begin{align}
   \Phi^{k_\perp;s}_m(\rho,\theta)
   =
   e^{\mathrm{i}m\theta} \,
   \left [
   \begin{array}{r}
      A^{k_\perp}\ J_{m}(k_\perp\rho) \\
      \frac{k_\perp}{\lambda_-} \, C^{k_\perp}\
      e^{\mathrm{i}\theta} \, J_{m+1}(k_\perp\rho) \\
      C^{k_\perp}\ J_{m}(k_\perp\rho) \\
      \frac{k_\perp}{\lambda_-} \, A^{k_\perp}\
      e^{\mathrm{i}\theta} \, J_{m+1}(k_\perp\rho)
   \end{array}
   \right ]
\label{k1_eigv2}
   ,
\end{align}
which has the correct boundary conditions irrespective of the value
of $m$. The fact that Eq.~\eqref{eq:K3caleq} allows for a solution
of the form~\eqref{k1_eigv2} is unlike the case of $(t,z)$ subspace
where the form of the solution~\eqref{eq:psi_2dc} is
mandatory~\cite{non_symmetric} in order to satisfy the
orthonormality conditions~\eqref{plus}.



\begin{thebibliography}{99}
\bibitem{disessa_bc}
   Positive- and negative-energy solutions, $\psi_{k}^{(\pm)}(x)$,
   are defined such that $\psi_{k}^{(\pm)}(x)$ vanish in the limit
   $t \rightarrow \mp \mathrm{i} \infty$,
   respectively~\cite{diSessa}. The Hankel functions of real argument
   $H^{(1,2)}_{\nu}(x)$ have the required properties. Hence,
   $\psi_{k}^{(+)}(x) \propto H^{(2)}_{\nu}(x)$ whereas
   $\psi_{k}^{(-)}(x) \propto H^{(1)}_{\nu}(x)$.
\bibitem{non_symmetric}
   The natural choice for the amplitude $\psi_{k_\eta}$ has $J_{\mathrm{i}k_\eta}$
   and $- e^{\eta} J_{\mathrm{i}k_\eta+1}$ as the upper and lower
   components, respectively. However, this choice results in a
   modified form of the orthonormality relations~\eqref{plus},
   which involves $\delta(k_\eta' - k_\eta -
   \mathrm{i})$. This suggests that the upper and lower components
   in $\psi$ must be ``symmetrized'' with respect to~$\eta$.
\bibitem{Kcomment}
   Since the gamma matrices anticommute, $\{\gamma^\mu,\,
   \gamma^\nu\}=0$ for $\mu \neq \nu$, it follows that a factor
   $\mathrm{i}\, \gamma^1\gamma^2$ is necessary in order to make the
   operators~$\mathcal{K}_1$ and $\mathcal{K}_1$ commute. For this
   to happen, the term containing the mass $M$ must be included in
   $\mathcal{K}_2$, and $\mathcal{K}_2$ gives the form of the
   Dirac spinor. If instead we use the factor
   $\mathrm{i}\, \gamma^0\gamma^3$, then the mass must be in
   $\mathcal{K}_1$, and the form of the Dirac spinor is given by solving the
   eigenvalue problem for $\mathcal{K}_1$.
\end{thebibliography}
\end{document}